\documentclass[a4paper,11pt]{article}
\pdfoutput=1 

\usepackage{jinstpub} 

\usepackage{lineno}
\usepackage{subcaption}

\title{\boldmath Prototype Design of a Timing and Fast Control system in
	the CBM Experiment}

\author[a]{V. Sidorenko,}
\author[b]{I. Fröhlich,}
\author[b]{W.F.J. Müller,}
\author[b]{D. Emschermann,}
\author[a]{S. Bähr,}
\author[b]{C. Sturm,}
\author[a]{J. Becker}

\affiliation[a]{Karlsruhe Institute of Technology, Institute for Information Processing Technologies\\Engesserstraße 5, 76131 Karlsruhe, Germany}
\affiliation[b]{GSI Helmholtz Centre for Heavy Ion Research,\\Planckstraße 1, 64291 Darmstadt, Germany}

\emailAdd{vladimir.sidorenko@kit.edu}

\abstract{The Compressed Baryonic Matter (CBM) experiment is designed to handle interaction rates of up to 10 MHz and up to 1 TB/s of raw data generated. With triggerless streaming data acquisition in the experiment and beam intensity fluctuations, it is expected that occasional data bursts will surpass bandwidth capabilities of the Data Acquisition System (DAQ) system. In order to preserve integrity of event data, the bandwidth of DAQ must be throttled in an organised way with minimum information loss. The Timing and Fast Control (TFC) system provides a latency-optimised datapath for throttling commands and distributes a system clock together with a global timestamp. This paper describes a prototype design of the system with focus on synchronisation and its evaluation.}

\keywords{Detector control systems (detector and experiment monitoring and slow-control systems, architecture, hardware, algorithms, databases); Control and monitor systems online}

\collaboration[c]{on behalf of CBM collaboration}

\proceeding{Topical Workshop on Electronics for Particle Physics\\
  2021}

\begin{document}
\maketitle
\flushbottom

\section{Introduction}
\label{sec:introduction}

The Compressed Baryonic Matter (CBM) experiment aims to study rare probes produced in heavy ion interactions. With the intended interaction rate of up to 10 MHz, the experiment will be equipped with self-triggered front-end electronics and a free-streaming data acquisition (DAQ) system. 1 TB/s of timestamped raw experimental data will be generated by the front-end electronics and preprocessed in parallel by a layer of 200 FPGA-based Common Readout Interface (CRI) boards. These boards also serve as a gateway for the data to a computing farm hosting the First-Level Event Selector (FLES) network, where online event reconstruction takes place.

Beam intensity fluctuations are expected in the experiment~\cite{Rost2019}. They are likely to cause localised congestion of the DAQ system. Partial loss of raw data leads to collecting corrupted events that cannot be reconstructed. A prior study demonstrated that collection of complete events in these conditions can be maximised using a centralised throttling mechanism~\cite{Gao2019}. The throttling decision is based on buffer and link occupancy in the DAQ network and is distributed to CRI boards to prevent uncontrolled data loss. A major prerequisite for throttling implementation is a low-latency control network with under 6 $\mu$s round-trip time~\cite{Gao2020}.

The current article presents a prototype design of a Timing and Fast Control (TFC) system. The system is meant to provide the framework for throttling implementation as well as sub-nanosecond precision clock and timestamp distribution to the experimental electronics. The paper starts with the high-level concept of the TFC network. Afterwards, it covers firmware architecture and time distribution mechanism. In the end, the evaluation setup is described and the results are presented.

\section{Concept of the Timing and Fast Control system}
\label{sec:concept}

The proposed TFC network has hierarchical structure with a single controller and focus on scalability. Network nodes are FPGA boards interconnected with bidirectional optical links.

Endpoint nodes collect and aggregate FIFO/network occupancy information. This information is then forwarded by a hierarchy of Submasters to a central TFC Master node. This node is the root of the system and is responsible for making the throttling decision. After that, a throttling command is broadcasted downstream to Endpoints.

\begin{figure}[h]
	\centering
	\includegraphics[width=.9\textwidth]{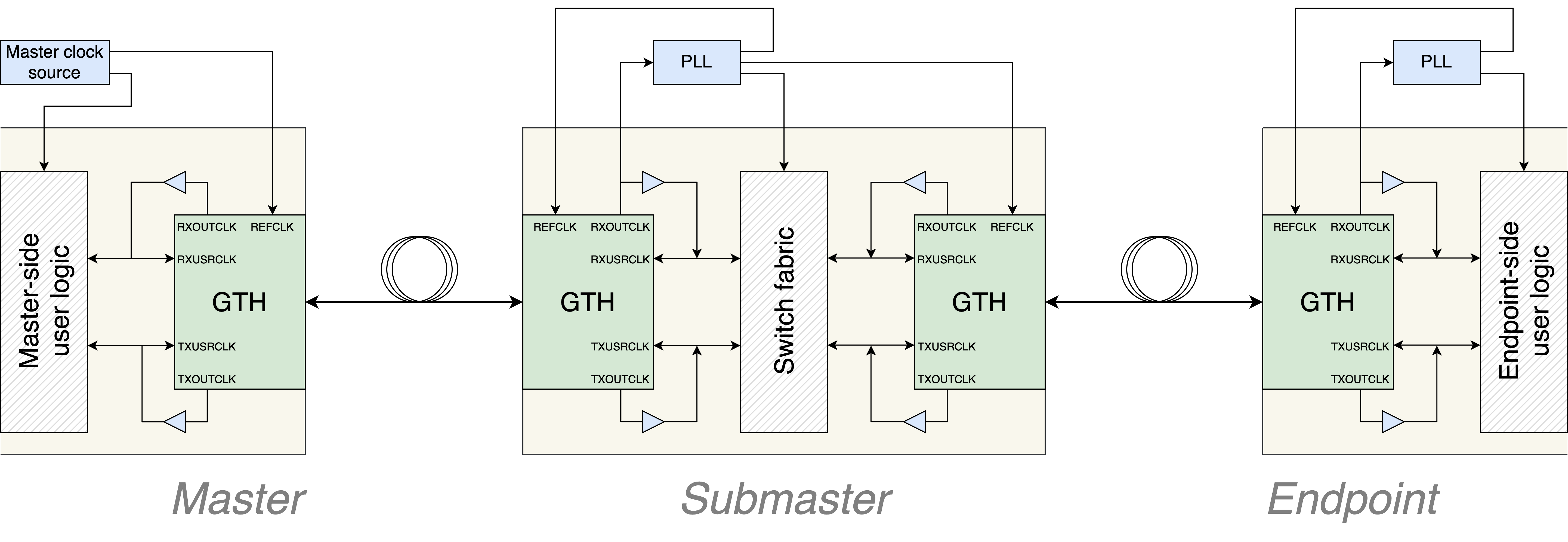}
  \captionsetup{justification=centering}
	\caption{\label{fig:cascaded_clocking} Cascased clock propagation scheme. Each subsequent node passes recovered clock through a jitter attenuator and reuses it for further transmission.}
\end{figure}

Another purpose of the TFC system is experiment-wide distribution of a clock and a 64-bit global timestamp. Cascaded clock recovery is used to propagate the clock from Master to Endpoints. In this scheme, a Gigabit Transceiver (GTH) in each Submaster node recovers the clock from an upstream link and reuses it for all downstream connections (Figure~\ref{fig:cascaded_clocking}). With the constraints of even network depth and deterministic downstream latency, global clock reaches all Endpoints with the same certain skew relative to the Master node.

\section{Firmware development}
\label{sec:development}

Firmware is developed in three variants, one for each TFC node role. Roles differ by the number and direction of the TFC interfaces. Master node only has a number of downstream connections, whereas Submaster boards also require one TFC link in the Master-side direction. From this perspective, Endpoints are the simplest nodes and need only one link to communicate with the TFC network.

\begin{figure}[h]
	\centering
	\includegraphics[width=.9\textwidth]{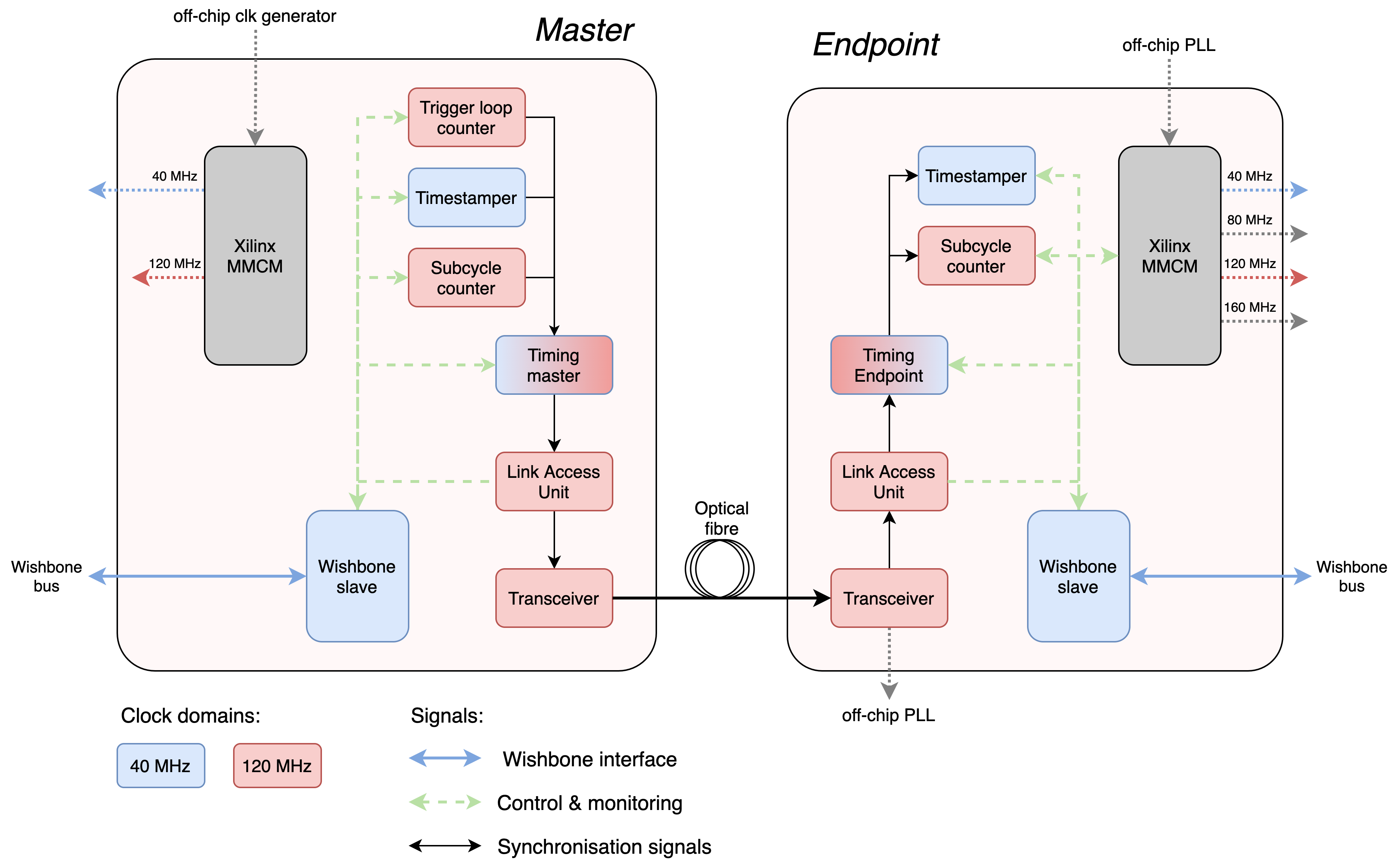}
  \caption{\label{fig:firmware_arch} Architecture of Master and Endpoint TFC cores.}
\end{figure}

To have a local notion of time, each TFC node runs a 64-bit timestamp counter at the base frequency of the experiment - 40 MHz. The counter is connected to a dedicated finite-state machine (FSM) that functions according to link direction. If the FSM serves a downstream link, it periodically transmits the local timestamp. In the case of an upstream link, the FSM receives incoming timestamps and applies a correction to the local counter if needed. Detailed architecture of the prototype firmware cores is shown in Figure~\ref{fig:firmware_arch}. Xilinx Multi-Mode Clock Manager (MMCM) primitives are used to synthesise various frequencies inside FPGA.

Protocol routines and link abstraction from the user logic are handled by a Link Access Unit (LAU). This module handles the necessary clock domain crossings with the transceiver and performs link initialisation and protocol check for link status monitoring (Figure~\ref{fig:tx_datapaths2}). In future implementations, LAU will also be responsible for shared link access for timing and fast control messages.

Various components of any variant of the TFC core are controlled over Wishbone bus. Provided that control software can access the same bus, the design allows to leverage the benefits of hardware/software co-design.

Major system clock frequency $f_{sys}$ in the experiment is 40 MHz. Data transport logic operates at a higher frequency $f_{tr}$ = 120 MHz. To ensure deterministic phase of $f_{sys}$ clock at the endpoint, a concept of subcycles was introduced. A counter runs at the higher of the frequencies and counts up to $\frac{f_{tr}}{f_{sys}}$, thus unambiguously labeling the edges relative to the slower clock. This way, alignment of the slower clock with any edge of the faster clock can be characterised by a value of this counter. Based on this value, an Endpoint node can recover the phase of its system clock.


\section{Implementation}
\label{sec:implementation}

Prototype firmware has been developed on the BNL-712 platform~\cite{bnl712}. These boards feature a Kintex Ultrascale FPGA, Si5345 jitter attenuator and 8 MiniPOD fiber optics modules that provide access to 48 bidirectional optical connections. 

Clock is recovered from incoming links using the built-in clock-data recovery mechanism of Xilinx GTH primitives. This clock must be cleaned with a PLL before it can be used by firmware components because of high jitter. In our case, it is passed through the onboard Si5345 jiter attenuator. Upon a TFC link loss, the internal clock switchover mechanism of Si5345 is used to smoothly transition to a local free-running reference. The phase shift rate of the output clock is limited at this point by the PLL bandwidth of 100 Hz. Thus, the clock remains uninterrupted at all times regardless of TFC link status.

For timestamp synchronisation, dedicated finite-state machines (FSM) were implemented in VHDL. The master-side FSM periodically serialises the current timestamp into 32-bit words and sends it to all downstream nodes and the endpoint FSM deserialises the timestamp and applies it to the local counter. The endpoint FSM also captures the subcycle at which the timestamp has arrived, and the system clock is aligned accordingly to ensure phase determinism.

\section{Prototype evaluation}
\label{sec:eval_setup}

\begin{figure}[h]
	\centering
	\begin{minipage}{.5\textwidth}
		\centering
		\includegraphics[width=.9\linewidth]{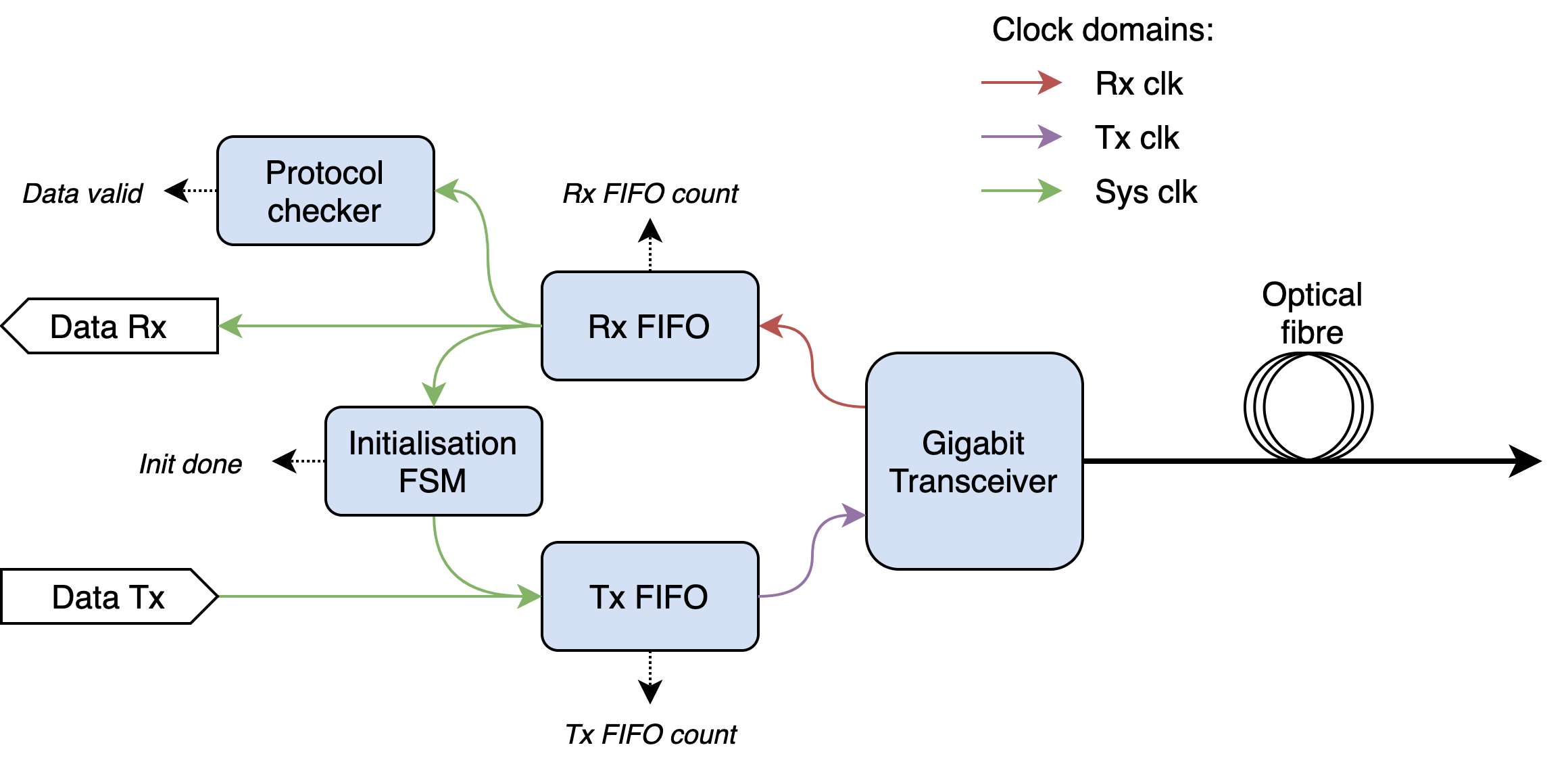}
		\captionof{figure}{Structure of Link Access Unit.}
		\label{fig:tx_datapaths2}
	\end{minipage}%
	\begin{minipage}{.5\textwidth}
		\centering
		\includegraphics[width=.9\linewidth]{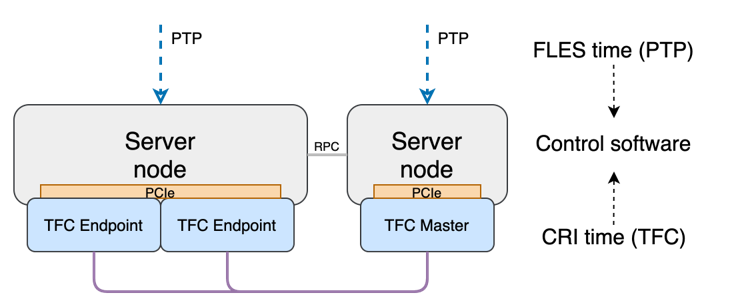}
    \captionsetup{justification=centering}
		\captionof{figure}{Test setup for evaluation of the prototype design.}
		\label{fig:test_setup}
	\end{minipage}
\end{figure}

In order to evaluate the design, a test setup has been integrated at GSI\footnote[1]{GSI Helmholtz Centre for Heavy Ion Research}, Germany (Figure~\ref{fig:test_setup}). Each BNL-712 has a slot for a timing mezzanine card (TMC) that adds an optical fibre interface to the board. One of such TMC cards is used on each TFC Endpoint node for upstream communication, since the MiniPOD connections are reserved for GBT links to transport experimental data.

As part of the test setup, one TFC Master and two Endpoints were mounted into server machines using PCIe. These machines were running Device Control Agent (DCA) software that can access FPGA register file. High-level test control was implemented as a Python script that uses DCA bindings and interacts with the server machines over Ethernet. To provide a common time reference for initial design evaluation, the server machines were synchronised with an off-the-shelf Precision Time Protocol (PTP) solution~\cite{ieee1588}.

\begin{figure}[h]
	\centering
	\begin{subfigure}{.5\textwidth}
		\centering
		\includegraphics[width=1\linewidth]{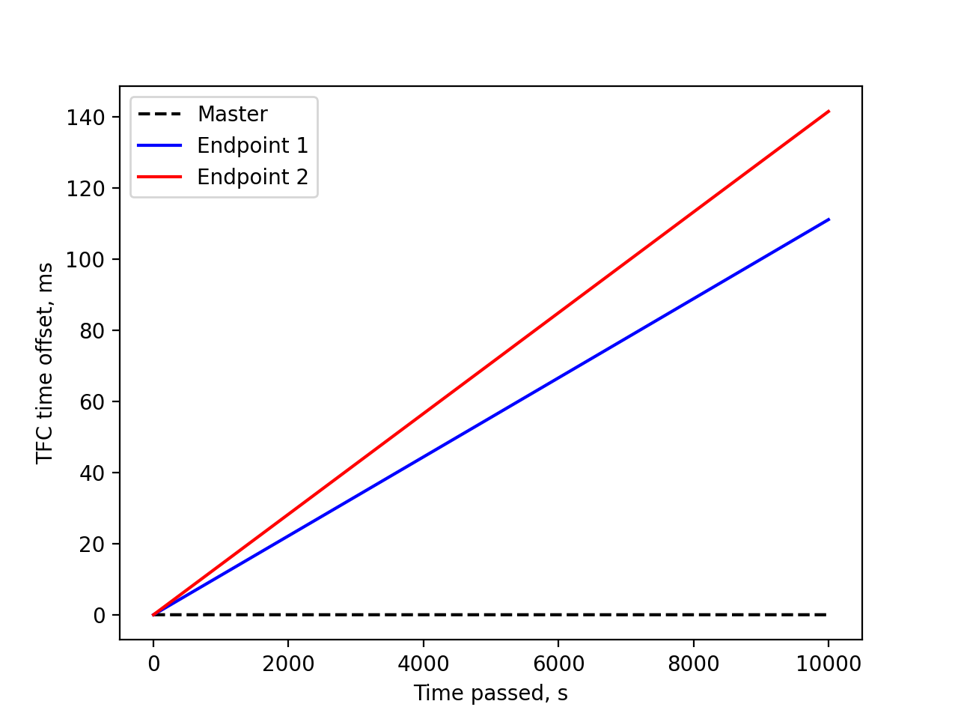}
		\caption{Asynchronous run.}
		\label{fig:test_async}
	\end{subfigure}%
	\begin{subfigure}{.5\textwidth}
		\centering
		\includegraphics[width=1\linewidth]{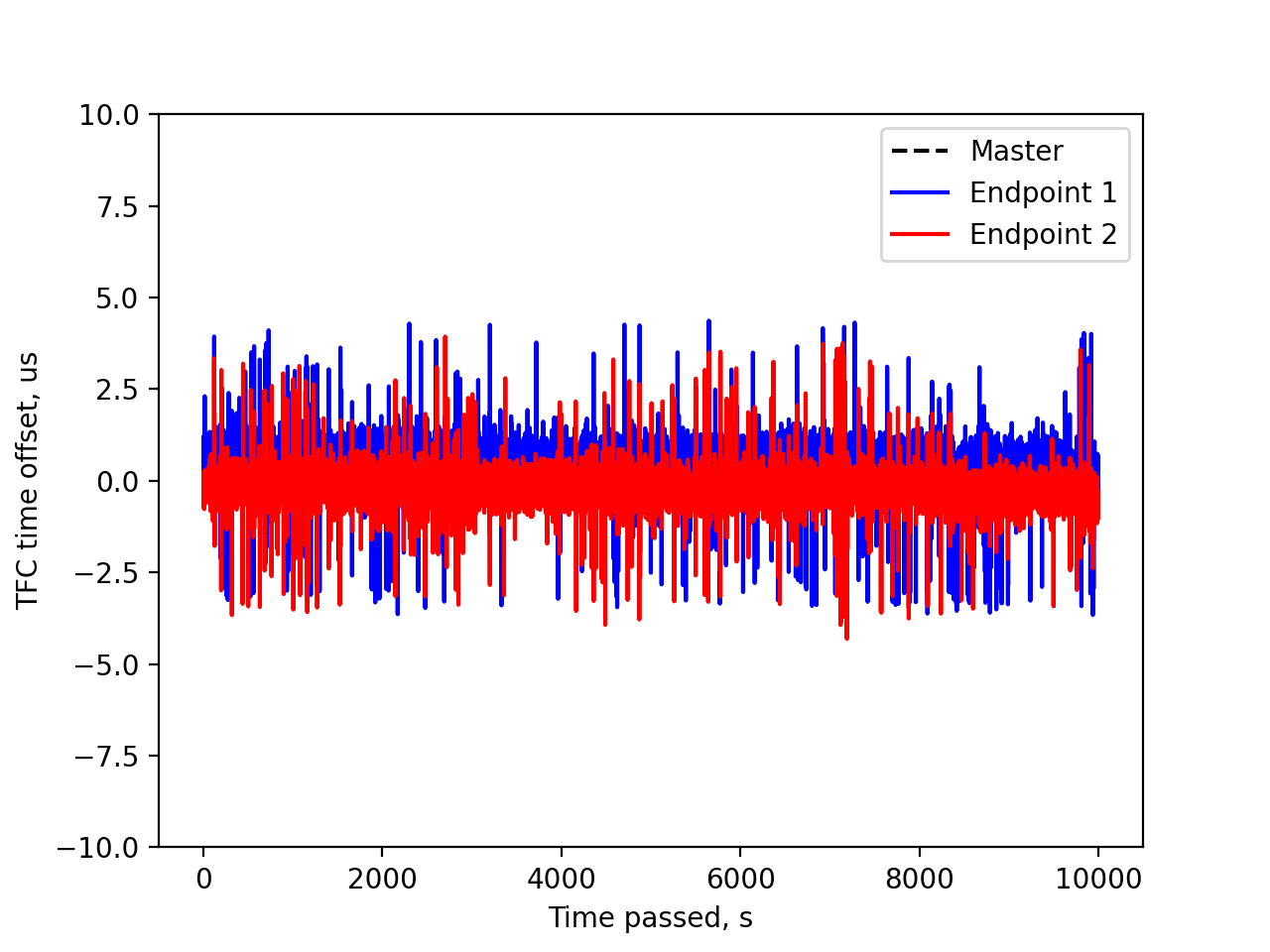}
		\caption{Synchronous run.}
		\label{fig:test_sync}
	\end{subfigure}
	\caption{Time drift between TFC nodes in asynchronous and synchronous modes.}
	\label{fig:lab_test}
\end{figure}

During the test, local timestamp drift was measured on each node with reference to the common PTP time. Local timestamps were recorded each second over a period of $10^4$ s. The test itself was run twice: first with clock and timestamp synchronisation disabled and the second time with synchronisation turned on. As a result, local time drifted off during the asynchronous run by nearly 30 ms between the Endpoints and both Endpoints drifted from the Master node by over 100 ms (Figure~\ref{fig:test_async}). With synchronisation on, no time drift has been observed and all three measured clocks remain synchronous with a constant phase offset (Figure~\ref{fig:test_sync}). Nevertheless, precision of the test is limited due to using PTP as a reference and uncertainty of timestamp readout latency.

In addition to the lab test, the prototype has been integrated into the experimental setup with 8 CRI Endpoints included. The setup has successfully passed the beam test and average collision rate of 1 MHz has been achieved.

\section{Conclusion and outlook}
\label{sec:conclusion}

As a result of the presented work, a prototype of the Timing and Fast Control system for the CBM experiment has been developed with focus on synchronisation functionality. It features system-wide clock distribution through cascaded clock recovery and periodic broadcast of the reference timestamp. To support future data throttling functionality, the prototype provides bidirectional optical links from Master to all Endpoints.

Clock distribution has been evaluated on a minimal lab setup as well as in a beam test and it has been demonstrated that there is no significant clock drift between network nodes present. As the further steps for evaluation, long-term stability and precision clock drift tests are planned.

Nonetheless, accuracy of time synchronisation in the current prototype design strongly depends on deterministic latency of downstream communication. Since data buffers are used in transceiver datapaths to handle clock domain crossing, this is a bottleneck that will be addresses in future work.


\begin{thebibliography}{99}

\bibitem{Rost2019}
  A. Rost, J. Adamczewski-Musch, T. Galatyuk, S. Linev, J. Pietraszko, M. Sapinski et al.,Performance of the CVD Diamond Based Beam Quality Monitoring System in the HADES Experiment at GSI, in Proceedings of International Particle Accelerator Conference, Melbourne, Australia, 2019 \url{doi:10.18429/JACoW-IPAC2019-WEPGW019}

\bibitem{Gao2019}
  X. Gao, D. Emschermann, J. Lehnert, W.F.J. Müller, Throttling Studies for the CBM Self-triggered Readout, in Proceedings of Topical Workshop on Electronics for Particle Physics, Santiago de Compostela, Spain, 2019 \url{doi:10.22323/1.370.0085}

\bibitem{Gao2020}
  X. Gao, D. Emschermann, J. Lehnert, W.F.J. Müller, Throttling strategies and optimization of the trigger-less streaming DAQ system in the CBM experiment, Nuclear Instruments and Methods in Physics Research Section A: Accelerators, Spectrometers, Detectors and Associated Equipment, Vol. 978, 2020, Art. 164442 \url{doi:10.1016/j.nima.2020.164442}

\bibitem{bnl712}
  K. Chen, H. Chen, J. Huang, F. Lanni, S. Tang and W. Wu, A Generic High Bandwidth Data Acquisition Card for Physics Experiments, in IEEE Transactions on Instrumentation and Measurement, vol. 69, no. 7, pp. 4569-4577, July 2020, \url{doi:10.1109/TIM.2019.2947972}

\bibitem{ieee1588}
	IEEE Standard for a Precision Clock Synchronization Protocol for Networked Measurement and Control Systems, in IEEE Std 1588-2019 (Revision ofIEEE Std 1588-2008) , vol., no., pp.1-499, 16 June 2020, \url{doi:10.1109/IEEESTD.2020.9120376}

\end{thebibliography}
\end{document}